\documentclass[twocolumn]{article}
\usepackage{blindtext}
\usepackage{multicol}
\usepackage{geometry}
\usepackage{graphicx}
\usepackage{mathpazo}

\usepackage[numbers,square]{natbib}
\usepackage[hyphens]{url}

\renewcommand\harvardurl[1]{\textbf{URL:} \url{#1}}
\usepackage{hyperref}

\usepackage{listings}
\usepackage{color}

\definecolor{mygray}{rgb}{0.4,0.4,0.4}
\definecolor{mygreen}{rgb}{0,0.8,0.6}
\definecolor{myorange}{rgb}{1.0,0.4,0}
\usepackage{tikz,xcolor,hyperref}

\definecolor{lime}{HTML}{A6CE39}
\DeclareRobustCommand{\orcidicon}{
	\begin{tikzpicture}
	\draw[lime, fill=lime] (0,0) 
	circle [radius=0.16] 
	node[white] {{\fontfamily{qag}\selectfont \tiny ID}};
	\draw[white, fill=white] (-0.0625,0.095) 
	circle [radius=0.007];
	\end{tikzpicture}
	\hspace{-2mm}
}
\foreach \x in {A, ..., Z}{\expandafter\xdef\csname orcid\x\endcsname{\noexpand\href{https://orcid.org/\csname orcidauthor\x\endcsname}
			{\noexpand\orcidicon}}
}

\title{Seamless GPU acceleration for C++ based physics with the\\Metal Shading Language on Apple's M series unified chips}

\usepackage{authblk}

\author[1]{Lars Gebraad}
\author[1]{Andreas Fichtner}

\affil[1]{Seismology and Wave Physics,\protect\\ETH Z\"urich}

\date{\today}

\begin{document}

\twocolumn[
  \begin{@twocolumnfalse}
    \maketitle

\begin{abstract}
The M series of chips produced by Apple have proven a capable and power-efficient alternative to mainstream Intel and AMD x86 processors for everyday tasks. Additionally, the unified design integrating the central processing and graphics processing unit, have allowed these M series chips to excel at many tasks with heavy graphical requirements without the need for a discrete graphical processing unit (GPU), and in some cases even outperforming discrete GPUs.

In this work, we show how the M series chips can be leveraged using the Metal Shading Language (MSL) to accelerate typical array operations in C++. More importantly, we show how the usage of MSL avoids the typical complexity of CUDA or OpenACC memory management, by allowing the central processing unit (CPU) and GPU to work in unified memory. We demonstrate how performant the M series chips are on standard one-dimensional and two-dimensional array operations such as array addition, SAXPY and finite difference stencils, with respect to serial and OpenMP accelerated CPU code. The reduced complexity of implementing MSL also allows us to accelerate an existing elastic wave equation solver (originally based on OpenMP accelerated C++) using MSL, with minimal effort, while retaining all CPU and OpenMP functionality.

The resulting performance gain of simulating the wave equation is near an order of magnitude for specific settings. This gain attained from using MSL is similar to other GPU-accelerated wave-propagation codes with respect to their CPU variants, but does not come at much increased programming complexity that prohibits the typical scientific programmer to leverage these accelerators. This result shows how unified processing units can be a valuable tool to seismologists and computational scientists in general, lowering the bar to writing performant codes that leverage modern GPUs.
\end{abstract}
  \end{@twocolumnfalse}
]

\section{Introduction}

Scientific computing has always been at the forefront of technological developments in computing. On the point of the latest ARM chips produced by Apple, the M series, the scientific community should act no different. This series of chips is a relatively new component of MacBooks produced by Apple. These are a ARM based chips, which means that its instruction set is fundamentally different from typical notebook, workstation and HPC suited processors from Intel or AMD. According to Apple, the usage of ARM chips in high performance notebooks and workstations has seen yielded increased performance and power efficiency\cite{apple2020}. 

\begin{figure}[t]
    \centering
    \includegraphics[width=\linewidth]{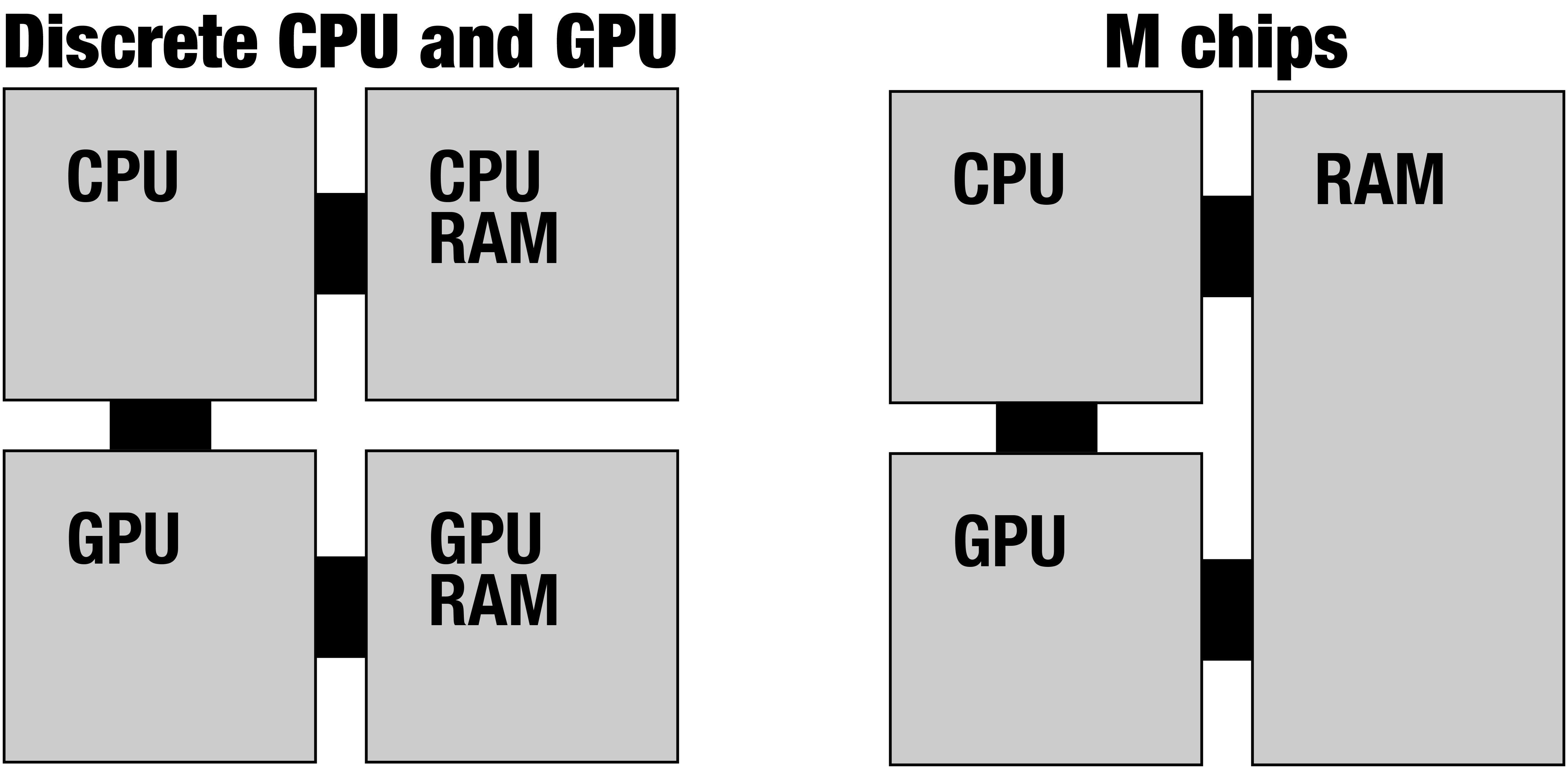}
 \caption{\label{fig:schematic}Schematic design of discrete processing unit systems versus the M series chips. On M systems, both processing units can talk to the same random access memory (RAM) without moving data locations}
\end{figure}

More importantly, however, is that the design of the M series chips are such that they combine the central and graphical processing units (CPU and GPU) onto a single chip, creating a 'Unified' Processing Unit, something atypical for modern systems. This means that both processing units can communicate with the same memory, simplifying many operations. The differences in design are schematically illustrated in Figure~\ref{fig:schematic}.

The Metal Shading Language (MSL), originally developed for the iPhone iOS operating system, is a programming language exclusive tailored to address GPU hardware present on Apple ARM chips in general. Although it has been in development for approximately a decade, it only saw its first stable release in 2019, and it's first non-mobile usage in 2020, when Apple released their first ARM-based MacBooks, fitted with M1 chips. Prior to this, MSL was only used to perform GPU operations on mobile devices. Although MSL is also able to control various GPUs from other manufacturers, we do not focus on systems with these GPUs in this work.

Although MSL is mostly focused on enabling graphics-oriented operations on Apple ARM GPUs, it also possesses functionality to instruct these chips for mathematical operations. This aspect of the M chip has, as it currently seems, not received much attention in the computational sciences, outside of the development of an MSL version of the popular TensorFlow software. This might be in part due to the fact that most documentation provided by Apple is focused on the Objective-C and Swift programming languages, typically favoured for developing general-purpose software on MacOS. The focus of these documentations is of course not the computational sciences, and as such, the exposure of the community to the MSL has so far been limited.

In this work, we illustrate the usage of MSL in general C++ array operations, and analyse how and when MSL provides benefit over running ``plain'' (multithreaded) CPU code. Special care will be given on how to enable the GPU operations in existing scientific C++ codes, specifically for numerical simulations of partial differential equations (PDEs) using finite differences. A case study focusing on accelerating elastic wave propagation in two dimensions using the M1 GPU illustrates the potential performance and ease of use of the M chip and unified processing units in general. All our simulations are run using single-precision decimal numbers (floats). Additionally, we provide a web portal that both links to our research codes as well as material helpful to the computational scientist to get started using MSL in C++ \cite{mslportal}.

\section{MSL execution and memory model}

MSL code itself compiles to instructions that are purely run on the GPU. These compiled functions are called shaders, or kernels. To orchestrate the execution of these kernels from the CPU, the instructions need to be fed from CPU code, i.e., any program we write in, e.g., C++. The communication of these instructions, as well as the scheduling of multiple operations and other ``steering'' tasks, are performed by the Metal Framework, an Objective-C library callable directly from C++ using `metal-cpp` \cite{metalcpp}. Performing operations on data with the Metal Framework on a GPU follows a set collection of steps, after \cite{metalmodel}:

\begin{enumerate}
    \itemsep0em 
    \item Create a command buffer and encoder. These objects respectively receive instructions (buffer), and encode them into machine language for the appropriate GPU (encoder);\\
    \item Place instructions and data addresses in encoder;\\
    \item Encode instructions with the encoder;\\
    \item Execute instructions.\\
\end{enumerate}

During the execution of the command buffer, the CPU can resume operation and synchronize with GPU execution at a later stage. Although one does need to instruct the GPU which data to use for these operations, the data itself does not need to be communicated to the GPU, by virtue of the unified design. 
The Metal Framework allows one to create this shared data in existing applications, and then simply get a standard C++ pointer to the underlying data such that it can be used in existing CPU code. This greatly simplifies exposing existing codes' array to new GPU operations, compared to using e.g. NVidia's CUDA\cite{cuda}. Trying to access the data from both the CPU and GPU simultaneously creates race conditions, and should be avoided.

      
    


Although encoding the commands and data in the buffer is more intricate than typical C++ operations, we provide a simple interface to these operations for both one-dimensional and two-dimensional data on our web portal.

\section{Basic array operations}

To introduce how performant the M GPU is with respect to (M) CPU configurations, we benchmark various "single instruction, multiple data", or SIMD, operations on both processing units for one-dimensional data. We test the performance of the SAXPY operation (Single-Precision $a \cdot x + y$) and a  3-point central differencing scheme. By testing these operations on arrays of different sizes, we can thoroughly demonstrate the overhead required for using both OpenMP and MSL configurations. To generate these numbers, we ran the operations repeatedly on a 2021 MacBook Pro equipped with the M1 Max 10 core CPU (of which 8 performance cores) and a 32 core GPU.

\begin{figure}[t]
    \centering
    \includegraphics[width=\linewidth]{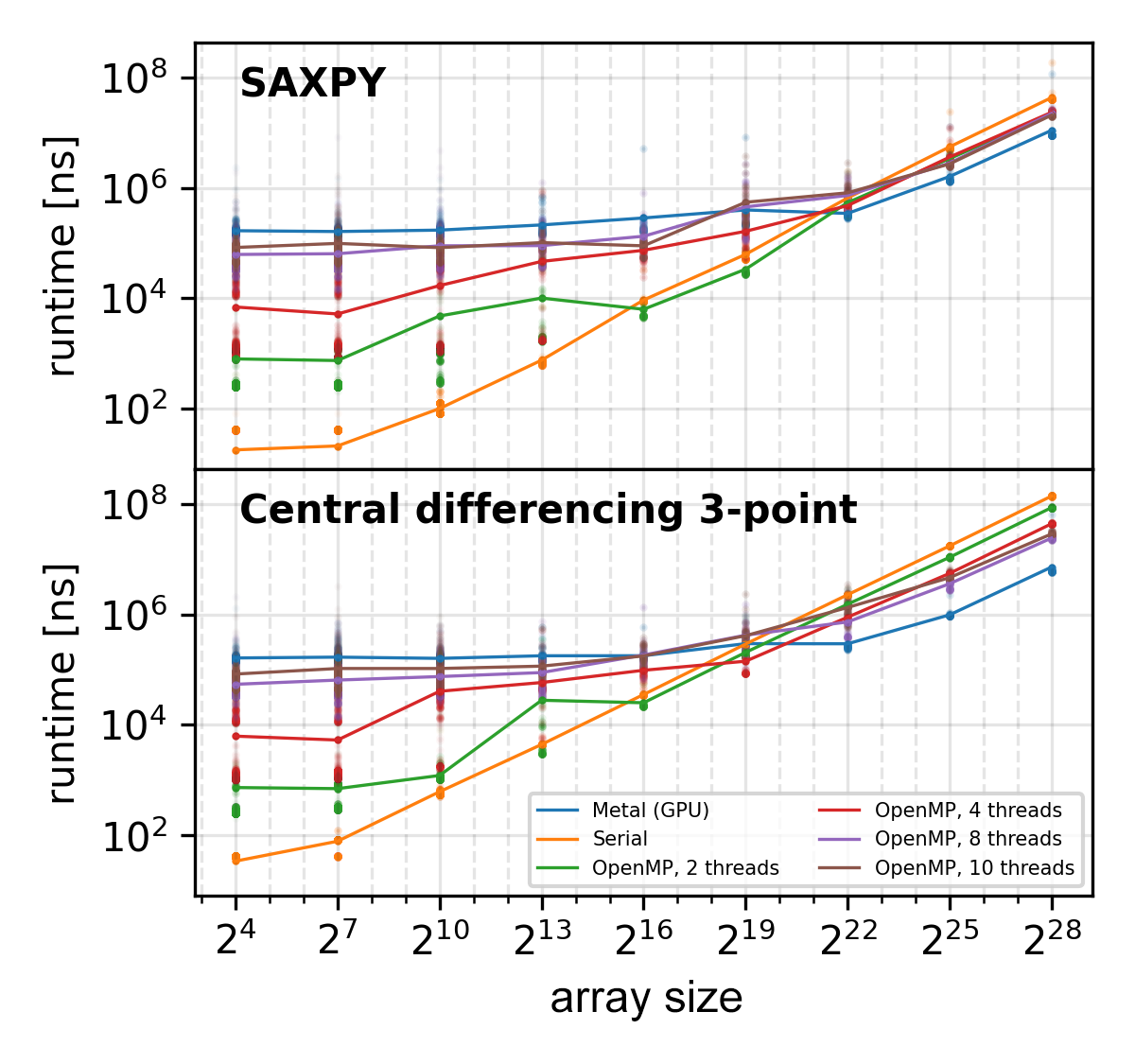}
 \caption{\label{fig:1d_ops}Runtime of one-dimensional operations on data of various sizes. Note that the median runtime is connected with the line, but all individual runs (i.e. multiple per data size per configuration) are indicates by circles, demonstrating variability of runtimes. Note that the runtime of OpenMP using 8 threads is always shorter than that of 10 threads, likely because the 2 performance cores of the M1 chip slow OpenMP scheduling down.}
\end{figure}

Figure~\ref{fig:1d_ops} shows how operations in GPU configurations, as well as operations in multithreaded CPU configuration, only provide benefit when the data is relatively large. Both array operations show fastest performance in serial mode up to an approximate data size of $2^{16}$ (= $65536$). The exact speed-up of using various configurations at largest data size is summarized in Table~\ref{tab:speedup}. These sepeed-ups demonstrate that even for simple operations, the usage of the GPU does accelerate one-dimensional array operations. The benefit of using the GPU for these relatively simple array operations in one dimension however only manifests itself at large data sizes, i.e. at $2^{22}$ elements and up.

\section{Multidimensional operations}

One staple of physical modelling are spatial derivatives, especially when solving PDEs (e.g. the wave equation) in two or three dimensions. MSL enables one to execute kernels specifically with two- or three-dimensional layouts, greatly simplifying the implementation of kernels that operate on two-dimensional or three-dimensional arrays.

\begin{figure}[t]
    \centering
    \includegraphics[width=1\linewidth]{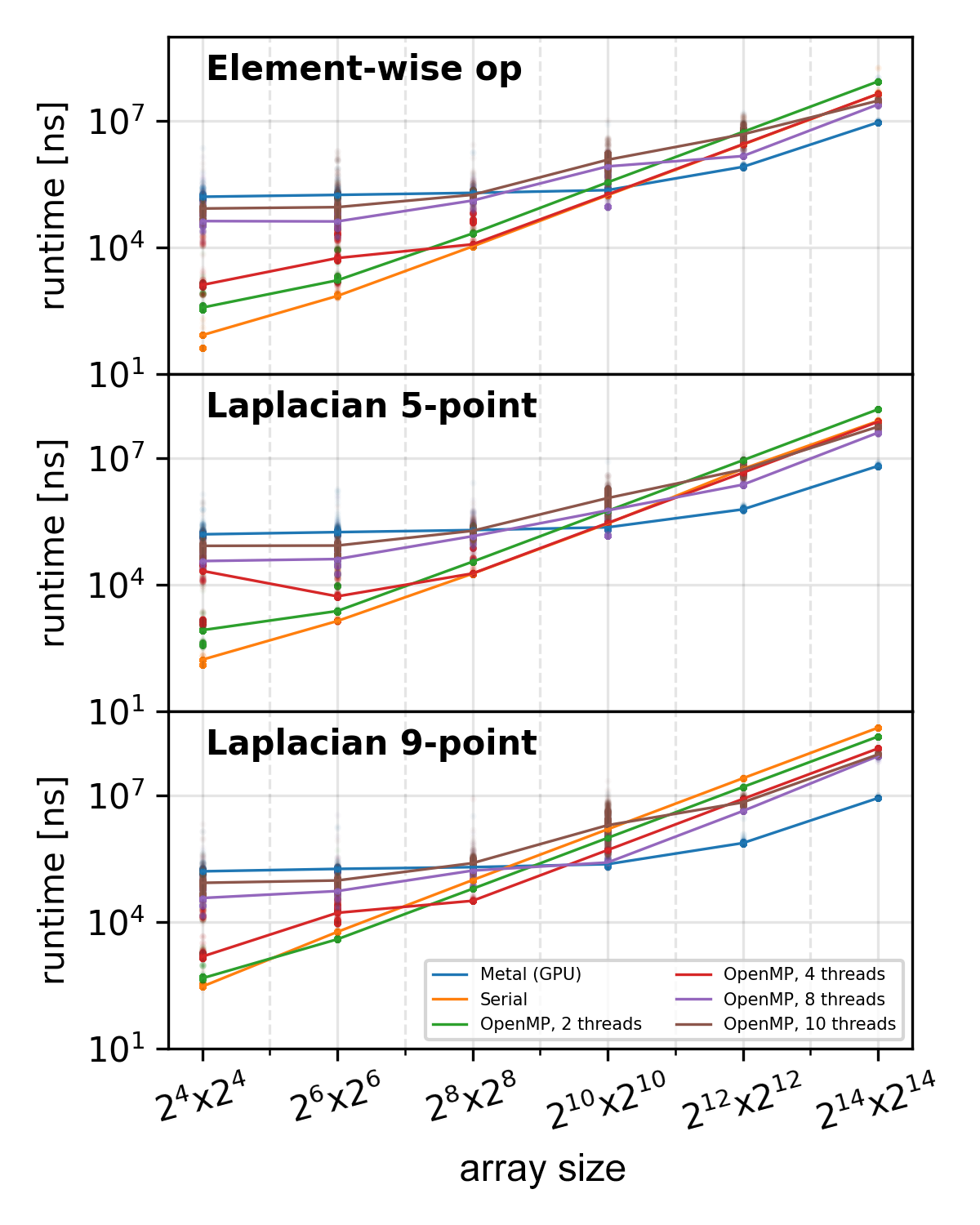}
    \caption{\label{fig:2d_ops}Runtime of two-dimensional operations on data of various sizes, always of equal width and height. As in Figure`\ref{fig:1d_ops}, separate runtimes are plotted using points, and the medians are connected by lines. The largest data-sizes show almost an order of magnitude speed-up between CPU (OpenMP 8 threads) and MSL GPU.}
\end{figure}

We detail the performance in CPU and GPU configurations by again performing benchmark of the operations on a 2021 M1 Max chip. The runtimes of these benchmarks are given in Figure~\ref{fig:2d_ops}. The operations again show dominance of the scheduling overhead (for both CPU and GPU) for multithreaded configurations at small data sizes compared to the serial configurations. The crossover point (for all operations) where GPU becomes the most performant configuration seems to be around a data size of $2^{10}\times2^{10}$ elements, close to the same number of effective elements for the crossover points in one dimension. 

As the number of instructions per operation increases (i.e. top to bottom in Figure~\ref{fig:2d_ops}), so does the benefit of both the multithreaded CPU and GPU configurations with respect to serial configuration. Table~\ref{tab:speedup} summarizes the speedup of the MSL, serial and optimal OpenMP configurations, showing how for large data sizes, MSL becomes more than an order of magnitude faster than OpenMP if the operation is complex.

\begin{table}[t]
    \centering
    \begin{tabular}{l|c|c|c}
        Operation & 8T vs 1T & MSL vs 1T & MSL vs 8T \\
        \hline
        1D SAXPY & 2.0x & 3.9x & 1.9x \\
        1D CD & 3.4x & 19x & 5.6x \\
        2D EW & 1.4x & 4.7x & 3.3x \\
        2D LA & 1.4x & 12x & 8.3x \\
        2D LA9p & 4.3x & 47x & 11x \\
    \end{tabular}
    \caption{Speed-up of various array operations at maximum tested array size. For the one-dimensional and two-dimensional operations respectively, these sizes $2^{28}$ and $2^{14}\times2^{14}$ elements. Speed-ups are calculated for OpenMP with 8 threads (8T) versus serial (1T), MSL versus serial, and MSL versus OpenMP with 8 threads. The operations summarized are those of the largest data sizes in Figures~\ref{fig:1d_ops} and \ref{fig:2d_ops}; single precision $a\cdot x + y$ (1D SAXPY), one-dimensional central differencing using a 3-point stencil (1D CD), two-dimensional element wise function (2D EW), two-dimensional Laplacian using a 5-point stencil (2D LA) and two-dimensional Laplacian using a 9-point stencil (2D LA9p).}
    \label{tab:speedup}
\end{table}

\section{Accelerating existing codes: elastic wave propagation on the M series GPU}

Possibly the biggest strength of using MSL is its drop-in capabilities. Because the CPU and GPU have unified memory, data created in this unified pool can be readily accessed by both processing units. Thus, MSL allows a user to readily modify existing C++ to be GPU-capable. 

Specifically, when creating an MSL buffer (i.e. an array that can be seen by the CPU and GPU), one can easily obtain a plain C++ pointer to the underlying data. This way, integrating MSL into an existing C++ application simply requires two additional lines per array: the declaration of the buffer and retrieving the raw C++ pointer (per Listing~1).

To demonstrate this capability, we took an existing two-dimensional elastic wave propagation code \cite{gebraad2022psvwave,Lars_psvWave_2022} based on Virieux's seminal paper \cite{virieux1986p}. This wave propagation code was developed to perform Full-Waveform Inversion (FWI), an approach to fit recorded vibrations to interior structure of materials. This method finds applications in for example seismology when imaging the entire Earth \cite{virieux2009overview, lei2020global,thrastarson2022data}, in non-destructive testing when imaging man-made structures\cite{nguyenNDT,kordjazi2020use}, and in medical tomography when imaging the human body \cite{guasch2020full,marty2021acoustoelastic}.

We implement the integration of the dynamic fields (material velocity in $x$ and $z$ direction, i.e. $v_x, v_z$, and vertical, horizontal and shear strain $\tau_{xx}, \tau_{zz}, \tau{xz}$) in MSL, but perform the rest of the operations required for FWI on CPU. These operations include recording the entire forward dynamical wavefields (for later use in the computation of sensitivity kernels), injecting point sources, recording wavefields at receivers, and the cross-correlation between forward and adjoint dynamical fields.

\begin{figure}[t]
    \centering
    \includegraphics[width=1\linewidth]{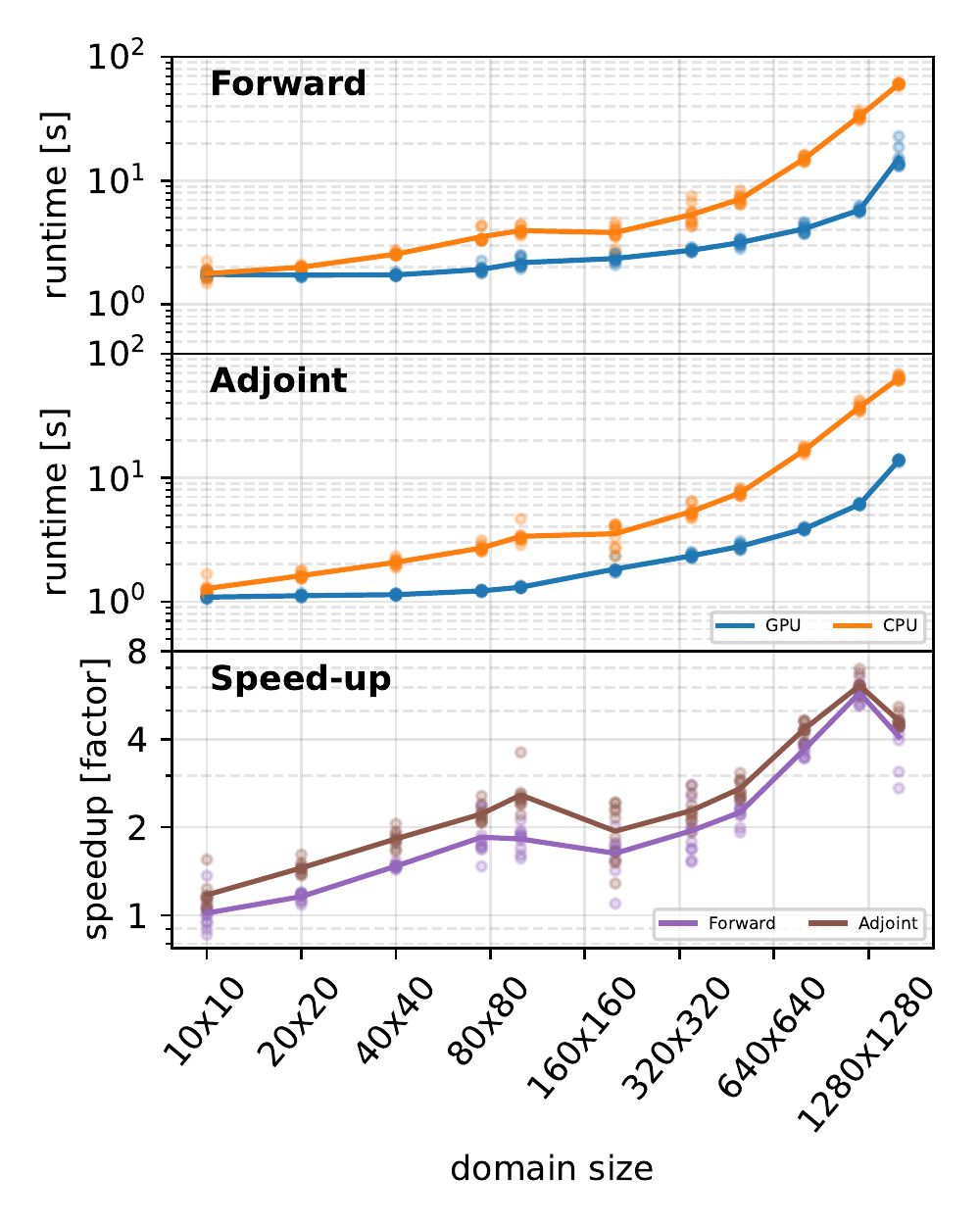}
    \caption{\label{fig:fwi}Runtimes of two-dimensional elastic wave propagation in media of various sizes for both OpenMP 8 threads configuration and MSL GPU configuration.}
\end{figure}

Figure~\ref{fig:fwi} demonstrates a surprising result with respect to our preceding results. At all domain sizes, the GPU configuration outperforms the 8-threaded OpenMP configuration. Where we were seeing cross-over points between the two configurations only at larger data sizes for simpler array operations, it now seems that the complexity of integrating wavefields means that MSL always outperforms OpenMP. 

As domain size grows, so does the speed-up of MSL with respect to OpenMP. The simulation code fails at domain sizes above approximately $2000\times2000$, as system RAM unpredictably runs out on our benchmarking machine. We were able to run a limited number of benchmarks at a domain size of $5000\times5000$, where the speed-up of MSL with respect to 8-threaded OpenMP was approximately a factor 10.

\section{Discussion}

Although the acceleration of computational physics by graphical processing units has long been acknowledged to yield performant codes, the complexity of implementation does create a barrier to its actual usage. The M chip series might herald the start of a paradigm-shift in computational sciences towards usage of unified chips and their programming languages. Although the M1 chip is not the first unified chip to be commercially available, nor is our implementation the first GPU-accelerated physical simulation, its ease of implementation does demonstrate the first implementation on a widely available chip using minimal effort. The upcoming release of the M2 chip promises increased unified memory bandwidth, thus further enlarging the potential for M series chips \cite{apple2022}.

Although for simple operations the speed-up of MSL with respect to OpenMP only becomes apparent at relatively large data sizes, in practical use for computational physics this threshold is much lower, due to the amount of instructions of the operations. This means that practically for our studied example (elastic wave propagation), it is typically worthwhile to use MSL over OpenMP. It is, however, not limited to our specific physics or numerical solver; these concepts of acceleration translate well to other PDEs solved with finite differences as well as to other numerical methods such as the finite element method\cite{rietmann2012forward,kiss2012parallel}.

\subsection{ThreadGroupSize dimensions}

 When using MSL shader functions, one needs to define in what shape the GPU traverses the operation. This access pattern is designated ThreadGroupSize in MSL, and is defined by a one-, two- or three-dimensional vector, depending on the input data. These vectors define in what pattern the multiple parallel cores on the GPU operate on the input data. 
 
 As an example, we show how we define the traversing of two-dimensional array operations for arrays with $n_x$ rows and $n_y$ columns. C++ arrays have linear memory layout and are indexed in two dimensions using the following linear index $i$:
\begin{equation}
    i = i_x \cdot n_y + i_y
\end{equation}
where $i_x$ is the index of the row, $i_y$ is the index of the column of the array. The ThreadGroupSize to launch two-dimensional kernels is defined by ($t_x$, $t_y$), where $t_x$ and $t_y$ indicate the dimensions of the tile of cores working on the data. The total amount of cores operating on the data in this tile is $t_x \cdot t_y$.

The usage of ThreadGroupSize is similar to CUDA scheduling of thread blocks. The access patterns of the data influence the performance of MSL and CUDA code alike. As an example, consider finite differencing schemes accessing neighboring elements. In these cases it is often beneficial to process multidimensional data in the order it is laid out in memory. Specifically, our two-dimensional data is laid out with strides of 1 element in the second dimension ($y$), and strides of $n_y$ in the first dimension ($x$). Therefore, we launch our kernels with ThreadGroupSizes of $(t_x, t_y)$, where $t_y >> t_x$. We find that this way our MSL kernels are the most performant. These optimal memory access patterns are well-known for general CPU and GPU programming, where the practice of optimizing access is typically known as memory coalescing \cite{memcoalesce,coalesce1994}. 

\subsection{Asynchronous operation}

As the CPU and GPU on the M series chips can operate asynchronously, there exists a potential further speed-up of array operations and computational tasks in general. We implemented this hybrid configuration for the two-dimensional elastic wave propagation. 

In this configuration, approximately half the workload is shifted from the GPU back to the CPU on the M chip by letting the CPU integrate the vertical velocity field and the shear strain field, while the other 3 fields are integrated by the GPU. As the strain fields depend on the velocity fields and vice-versa, synchronization between the GPU and CPU is performed twice per time-step.

The results, however, are disappointing. For any configuration, the runtime of the hybrid configuration is approximately half that of the CPU configuration. This is because the M1 GPU significantly outperforms the M1 CPU for the elastic wave propagation, and spends most of the time waiting for synchronization. To actually attain a practical speed-up, the workload needs to be divided proportionally to the performance of both processing units, i.e. a larger part of the compute task needs to be allocated to the GPU. This was not implemented for our work.

\section{Acknowledgements}

We would like to thank the various members of the Seismology and Wave Physics Group for providing valuable feedback on and accommodating a proving ground for our MSL implementation.

\bibliography{references}

\end{document}